\documentclass{webofc}

\usepackage[varg]{txfonts}   
\usepackage{hyperref}
\usepackage{url}
\hypersetup{colorlinks=false,citecolor=blue,urlcolor=blue,linkcolor=blue}
\usepackage{lineno}
\begin{document}

%
\title{EPJ Featured Talk: First direct measurement of radial flow in heavy-ion collisions with ALICE}
%
%

\author{\firstname{Swati} \lastname{Saha}\inst{1,2}\fnsep\thanks{\email{swati.saha@niser.ac.in}} for the ALICE Collaboration 
}

\institute{School of Physical Sciences, National Institute of Science Education and Research, Jatni-752050, Odisha, India
\and
          Homi Bhabha National Institute, Training School Complex, Anushaktinagar, Mumbai-400094, Maharastra, India
          }

\abstract{This work presents measurements of the transverse-momentum-dependent observable $v_{0}(p_\mathrm{T})$ as a novel probe of radial expansion dynamics in Pb$-$Pb collisions at $\sqrt{s_\mathrm{NN}} = 5.02$ TeV with the ALICE detector. Results are reported for inclusive charged hadrons, pions, kaons, and protons across centrality intervals, using a pseudorapidity gap to suppress short-range nonflow correlations. At low $p_\mathrm{T}$, a clear mass ordering is observed, consistent with hydrodynamic expectations. For $p_\mathrm{T} > 3$ GeV/$c$, protons exhibit larger $v_{0}(p_\mathrm{T})$ than pions and kaons, in line with quark recombination models. These results demonstrate the sensitivity of $v_{0}(p_\mathrm{T})$ to collective expansion and hadronization dynamics in the quark--gluon plasma.}

\maketitle

\vspace*{-0.8cm}   

\section{Introduction}
\label{intro}
Heavy-ion collisions at the LHC and RHIC create a short-lived quark--gluon plasma (QGP), where quarks and gluons are deconfined under extreme conditions~\cite{Shuryak:1978ij, Cleymans:1985wb}. The QGP exhibits collective behavior well described by relativistic hydrodynamics~\cite{Ollitrault:2007du}, with its expansion comprising anisotropic flow, driven by differences in the pressure gradients due to the initial spatial asymmetries of the collision zone, and radial flow, which arises from the overall pressure pushing matter outward. 

While anisotropic flow is extensively studied via azimuthal correlations~\cite{ALICE:flow_h, ALICE:flow_pid}, direct radial flow measurements remain limited. Traditional methods, such as Boltzmann–Gibbs blast-wave fits to transverse momentum ($p_\mathrm{T}$) spectra, yield a single parameter per centrality but cannot resolve its $p_\mathrm{T}$ dependence~\cite{Schnedermann:1993ws, ALICE:blast}.

The recently introduced $v_{0}(p_\mathrm{T})$ observable enables $p_\mathrm{T}$-differential studies of radial flow, quantifying the long-range $p_\mathrm{T}$ correlations arising from the collective behavior of the medium. In (3+1)D hydrodynamic simulations, $v_{0}(p_\mathrm{T})$ reveals mass ordering at low $p_\mathrm{T}$, analogous to $v_{2}(p_\mathrm{T})$, reflecting the stronger boost heavier particles receive from collective expansion~\cite{Theory:1, Theory:2}. The observable is found to be sensitive to the QCD equation of state and bulk viscosity, but largely insensitive to shear viscosity, offering a complementary constraint on medium properties~\cite{Theory:1, Theory:2}.

This article reports on ALICE measurements of $v_{0}(p_\mathrm{T})$ for inclusive charged particles, pions, kaons, and protons in Pb$-$Pb collisions at $\sqrt{s_\mathrm{NN}} = 5.02$~TeV across centrality intervals.

\section{Observable}
The observable $v_{0}(p_\mathrm{T})$ in Eq.~\ref{eq:formula1} quantifies the normalized covariance between fluctuations of the event-wise mean transverse momentum ($[p_\mathrm{T}]$) and the fraction of particle yield ($f(p_\mathrm{T})$) in individual $p_\mathrm{T}$ bins. These quantities are calculated in separated pseudorapidity ($\eta$) windows (referred as $A$ and $B$, with a gap of $\Delta\eta$) to suppress nonflow contributions~\cite{Theory:1, Theory:2} as follows:

 \begin{equation}
\label{eq:formula1}
v_{0}(p_\mathrm{T}) = \frac{\langle f_{A}(p_\mathrm{T})[p_\mathrm{T}]_{B}\rangle - \langle f_{A}(p_\mathrm{T})\rangle\langle [p_\mathrm{T}]_{B}\rangle}{\langle f_{A}(p_\mathrm{T})\rangle\sigma_{[p_\mathrm{T}]}}.
\end{equation}
Here, angular brackets $\langle ...\rangle$ denote averaging over events, and $\sigma_{[p_\mathrm{T}]}$ denotes the standard deviation of event-by-event $[p_\mathrm{T}]$ fluctuations, given by
\begin{equation}
\label{eq:formula2}
\sigma_{[p_\mathrm{T}]}=\sqrt{\langle [p_\mathrm{T}]_{A}[p_\mathrm{T}]_{B}\rangle - \langle [p_\mathrm{T}]_{A}\rangle\langle [p_\mathrm{T}]_{B}\rangle}.
\end{equation}

\section{Analysis details}
This analysis is based on Pb--Pb collision data collected by the ALICE detector at the LHC during the 2018 data-taking period. The ALICE detector setup and performance are documented in Refs.~\cite{ALICE:Exp1}. Minimum-bias events are triggered by coincident signals in the V0 detectors (V0A and V0C)~\cite{ALICE:Exp2} covering pseudorapidity ranges $2.8<\eta<5.1$ and $-3.7<\eta<-1.7$, respectively. Events are filtered to ensure the primary vertex lies within $\pm10$~cm of the nominal interaction point along the beam axis, and pileup events with multiple primary vertices are rejected. The centrality of the events is determined from the V0 amplitude distribution. More details of the event and track selection, particle identification, and estimation of systematic uncertainties can be found in Ref.~\cite{ALICE:paper}.

\section{Results and discussions}
\begin{figure}[h]
\centering
\includegraphics[width=0.65\textwidth]{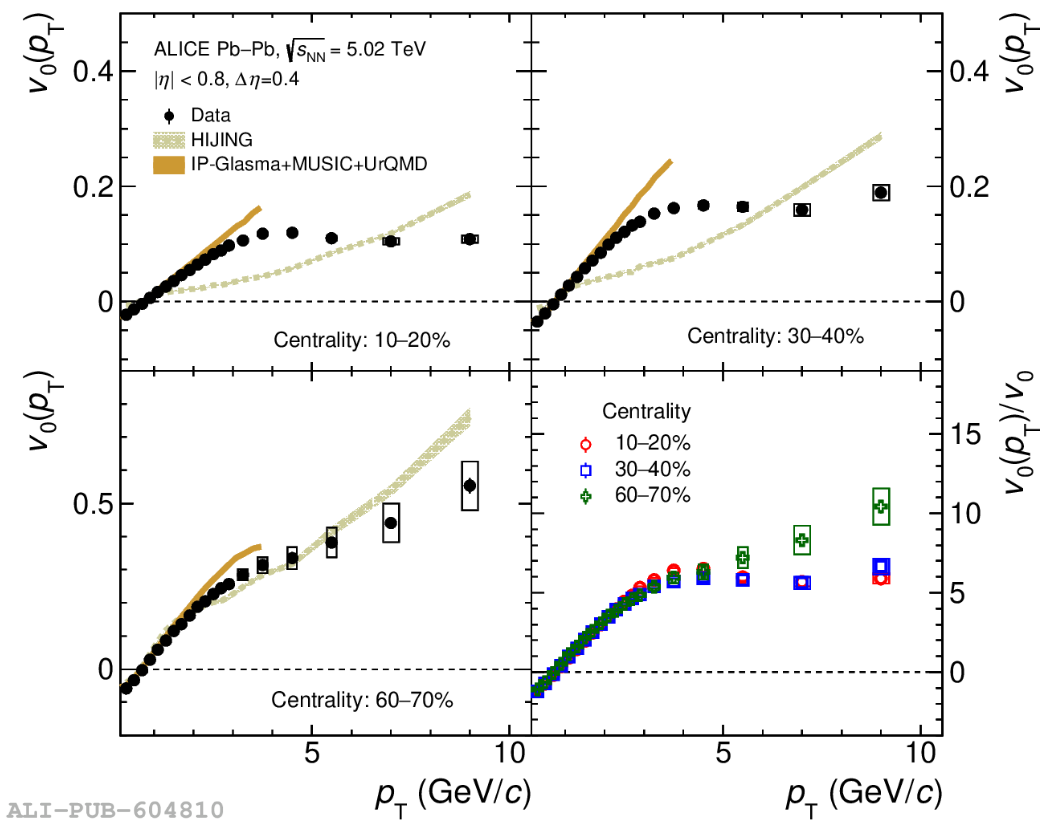}
\caption{\footnotesize The distributions of $v_0(p_\mathrm{T})$ for charged particles in Pb--Pb collisions at $\sqrt{s_{\mathrm{NN}}} = 5.02$~TeV, shown for central (10--20\%), semicentral (30--40\%), and peripheral (60--70\%) collisions. Comparisons with HIJING (dashed) and IP-Glasma+MUSIC+UrQMD (solid) models are shown. The scaled ratio $v_0(p_\mathrm{T})/v_0$ appears in the bottom right panel. Error bars (boxes) show statistical (systematic) uncertainties.}
\label{fig:had}
\vspace*{-0.7cm}  
\end{figure}

Figure~\ref{fig:had} presents the $p_\mathrm{T}$ dependence of $v_{0}(p_\mathrm{T})$ for inclusive charged particles in three representative centrality intervals: central (10--20\%), semicentral (30--40\%), and peripheral (60--70\%). In the low-$p_\mathrm{T}$ region ($p_\mathrm{T} < 0.8$~GeV/$c$), $v_{0}(p_\mathrm{T})$ is negative across all centralities. This behavior reflects an anti-correlation between event-by-event fluctuations of the mean $p_\mathrm{T}$ and particle production at low $p_\mathrm{T}$, consistent with earlier studies~\cite{Theory:1, Theory:2}. The negative sign arises from the construction of the observable: events with upward fluctuations in the mean $p_\mathrm{T}$ (i.e., a higher-than-average mean $p_\mathrm{T}$) tend to have more high-$p_\mathrm{T}$ particles and fewer low-$p_\mathrm{T}$ ones, while downward fluctuations do the opposite, thereby shaping the sign of $v_{0}(p_\mathrm{T})$. As $p_\mathrm{T}$ increases, the observable rises approximately linearly up to about 4.0~GeV/$c$, with the slope becoming steeper from central to peripheral collisions. This centrality dependent trend of increasing slope from central to peripheral collisions is consistent with increasing event-by-event fluctuations of $[p_\mathrm{T}]$, which directly correspond to fluctuations in radial flow~\cite{blast}. Beyond 4.0~GeV/$c$, the linear rise of $v_{0}(p_\mathrm{T})$ starts to deviate in central and semicentral collisions, where the slope decreases. In peripheral collisions, this deviation is less pronounced, likely reflecting differences in the relative contributions of hard and soft processes at high $p_\mathrm{T}$.

The IP-Glasma+MUSIC+UrQMD hydrodynamic model accurately describes the experimental data up to $p_\mathrm{T} \approx 2.0$~GeV/$c$ for all centrality classes. This framework combines three key elements: IP-Glasma initial conditions~\cite{IPglasma} describing the early-time energy density right after the collisions, MUSIC~\cite{MUSIC} for viscous hydrodynamic evolution, and UrQMD~\cite{UrQMD} for late-stage hadronic interactions. The model's success in reproducing various ALICE measurements---including particle yields, mean $p_\mathrm{T}$, and flow coefficients---stems from its incorporation of temperature-dependent shear ($\eta/s$) and bulk ($\zeta/s$) viscosities~\cite{Hydro1}. In contrast, the HIJING model~\cite{HIJING}, which accounts for mini-jet production and nuclear effects but doesn't include collective flow, fails to describe central and semicentral collision data where collective expansion dominates. However, it shows reasonable agreement with peripheral collision data even at higher $p_\mathrm{T}$, indicating that jet production and hard scattering processes become increasingly important in these smaller systems. The panel (d) of Fig.~\ref{fig:had} presents the $p_\mathrm{T}$ dependence of the scaled observable $v_0(p_\mathrm{T})/v_0$ for different centrality intervals. Across all centralities, the scaling behavior observed up to about 5~GeV/$c$ is consistent with hydrodynamic expectations~\cite{Theory:1,Theory:2}. However, above 5~GeV/c, deviations from this scaling appear, especially in peripheral collisions, reflecting increasing contributions from non-collective effects.

\begin{figure}
\centering
\includegraphics[width=0.8\textwidth]{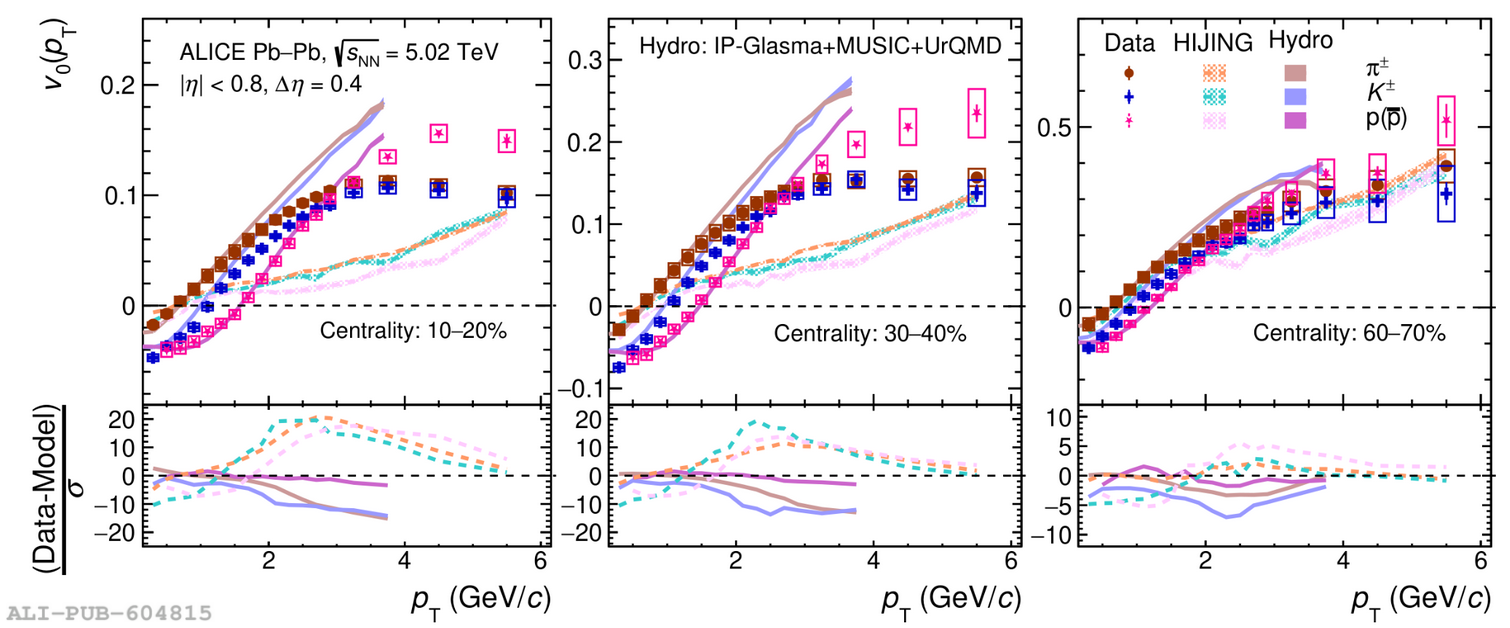}
\caption{\footnotesize The distributions of $v_0(p_\mathrm{T})$ for identified particles ($\pi^\pm$, $K^\pm$, $p/\bar{p}$) in Pb--Pb collisions at $\sqrt{s_{\mathrm{NN}}} = 5.02$~TeV, shown for central (10--20\%), semicentral (30--40\%), and peripheral (60--70\%) collisions. Data are compared to HIJING and IP-Glasma+MUSIC+UrQMD predictions. Statistical (systematic) uncertainties appear as vertical bars (boxes). Bottom panels display normalized residuals (Data--Model)/$\sigma$.}
\label{fig:piKp}
\vspace*{-0.7cm}  
\end{figure}

Figure~\ref{fig:piKp} presents $v_0(p_\mathrm{T})$ for identified hadrons ($\pi^\pm$, $K^\pm$, $p/\bar{p}$) measured up to 6~GeV/$c$ across three centrality classes. The observed $p_\mathrm{T}$ dependence is similar to the trend seen for inclusive charged hadrons. It also reveals two distinct phenomena: below 3~GeV/$c$, a clear mass ordering emerges ($v_0(p_\mathrm{T})^{\pi} > v_0(p_\mathrm{T})^\mathrm{K} > v_0(p_\mathrm{T})^{\mathrm{p}}$) consistent with hydrodynamic predictions~\cite{Theory:1}, while above this threshold, protons exhibit significantly larger $v_0(p_\mathrm{T})$ values than both pions and kaons. This is similar to the baryon-meson splitting observed in anisotropic flow coefficients~\cite{ALICE:flow_pid}, and suggests quark recombination as the dominant particle production mechanism at intermediate $p_\mathrm{T}$. The splitting is strongest in central (10--20\%) and negligible in peripheral (60–70\%) collisions, which may reflect system-size effects---with larger, denser systems favoring recombination and smaller ones being dominated by fragmentation~\cite{syssize}.

The IP-Glasma+MUSIC+UrQMD framework successfully reproduces the mass-ordered hierarchy at low $p_\mathrm{T}$, with quantitative agreement extending to approximately 3 GeV/$c$ for protons, 2 GeV/$c$ for pions, and 1.5 GeV/$c$ for kaons. In contrast, the HIJING model fails to capture both the low-$p_\mathrm{T}$ mass ordering (in central or semicentral collisions) and the high-$p_\mathrm{T}$ baryon-meson splitting, though it describes the data qualitatively in peripheral collisions, where non-collective processes dominate. 

\section{Summary}
In summary, the first measurement of the $v_0(p_\mathrm{T})$ observable, a sensitive probe of radial flow in heavy-ion collisions, is presented for inclusive charged hadrons, pions, kaons, and protons. The measured $p_\mathrm{T}$ dependence follows trends predicted by hydrodynamic model calculations. For identified particles, the results exhibit clear mass ordering at low $p_\mathrm{T}$ and baryon–meson separation at intermediate $p_\mathrm{T}$, consistent with hydrodynamic expansion and quark recombination~\cite{coalescence}. Hydrodynamic models successfully describe the low-$p_\mathrm{T}$ region, whereas deviations at high $p_\mathrm{T}$ point to an increasing role of non-collective processes. Overall, $v_0(p_\mathrm{T})$ provides complementary and novel insight into parton–medium interactions and hadronization dynamics, and its application to identified light hadrons paves the way for future studies involving heavy-flavor and rare particles.


%

\end{document}